\documentclass{iau_FM}
\usepackage{graphicx}

\def\la{\mathrel{\mathchoice {\vcenter{\offinterlineskip\halign{\hfil
$\displaystyle##$\hfil\cr<\cr\sim\cr}}}
{\vcenter{\offinterlineskip\halign{\hfil$\textstyle##$\hfil\cr
<\cr\sim\cr}}}
{\vcenter{\offinterlineskip\halign{\hfil$\scriptstyle##$\hfil\cr
<\cr\sim\cr}}}
{\vcenter{\offinterlineskip\halign{\hfil$\scriptscriptstyle##$\hfil\cr
<\cr\sim\cr}}}}}

\title[FM16.~~Clusters rich in RSGs] 
{Clusters rich in red supergiants}

\author[Ignacio Negueruela]   
{Ignacio Negueruela$^1$
}

\affiliation{$^1$Departamento de F\'{\i}sica, Ingenier\'{\i}a de Sistemas y
  Teor\'{\i}a de la Se\~{n}al\\ Escuela Polit\'ecnica Superior, Universidad de Alicante\\ Carretera de San Vicente del Raspeig s/n,  E03690, San Vicente del Raspeig, Alicante, Spain\\ email: {\tt ignacio.negueruela@ua.es}}

\pubyear{2016}
\jname{Astronomy in Focus, XXIXB, \textrm{Focus Meeting 16}} 
\editors{Ben Davies, ed.}
\begin{document}

\maketitle

\begin{abstract}
In the past few years, several clusters containing large numbers of red supergiants have been discovered. These clusters are amongst the most massive young clusters known in the Milky Way, with stellar masses reaching a few  $10^{4}\:$M$_{\odot}$. They have provided us, for the first time, with large homogeneous samples of red supergiants of a given age. These large populations make them, despite heavy extinction along their sightlines, powerful laboratories to understand the evolutionary status of red supergiants. While some of the clusters, such as the eponymous RSGC1, are so obscured that their members are only observable in the near-IR, some of them are easily accessible, allowing for an excellent characterisation of cluster and stellar properties. The information gleaned so far from these clusters gives strong support to the idea that late-M type supergiants represent a separate class, characterised by very heavy mass loss. It also shows that the spectral-type distribution of red supergiants in the Milky Way is very strongly peaked towards M1, while suggesting a correlation between spectral type and evolutionary stage.
\keywords{open cluster and associations: general, stars: massive, supergiants, evolution}
\end{abstract}

\firstsection 
\section{Introduction}

Young open clusters are natural laboratories for the study of high-mass stellar evolution. Red supergiants (RSGs) are believed to represent (part of) the He core burning phase of stars between~$\sim8$ and~$\sim25\:$M$_{\odot}$ and thus are found in clusters with ages between $\sim7$ and~$\sim30\:$Ma. These stars spend $\la10$\% of their lifetimes in the RSG phase, and so a cluster must have a moderately high number of massive stars if we want to have a reasonable chance of finding a RSG within its population. 
Simulations by Clark \etal\ (\cite[2009]{clark09a}) suggest that a cluster mass of $\sim10^{4}\:$M$_{\odot}$ is needed to have a high probability of finding more than one RSG. Nature seems to be a bit more generous, and at least two clusters with masses not quite reaching  $10^{4}\:$M$_{\odot}$ (NGC~884, NGC~7419) have five RSGs. Conversely, and in agreement with the simulations, some other clusters of comparable mass do not contain any RSG (e.g. NGC~869; in truth, assigning the membership of AD~Per to NGC~869, NGC~884, or none of them is a matter of taste, demonstrating from the beginning the weakness of number counts).

To understand RSGs, a combination of detailed studies of bright sources and population studies of large samples is needed. In addition, large samples of RSGs deliver information on the physics of stellar interiors and the validity of evolutionary models. For example, the ratio of red to blue supergiants in a population is a powerful test of evolutionary timescales. This ratio seems to increase with decreasing metallicity, and correlate strongly with the fraction of Be stars (\cite[Meynet \etal\ 2007; 2011]{meynet07,meynet11}). The ratio of Wolf-Rayet stars to RSGs also varies from galaxy to galaxy, again correlated to metallicity, and perhaps reflecting the impact of stellar winds and/or binarity (\cite[Massey 2003]{massey03}).  

Since the typical young cluster in the solar neighbourhood has a mass rarely exceeding 2 or $3\times10^{3}\:$M$_{\odot}$, the number of clusters with more than two RSGs known in the Galaxy is very small. Characterisation of RSGs thus requires alternative approaches, such as the observation of large samples in a given galaxy. It is possible to generate samples by adding up data in the literature for many small clusters (\cite[e.g. Eggenberger \etal\ 2002]{eggenberger2002}), but this procedure is subject to strong uncertainties due to very inhomogeneous sampling and membership assignment. A magnitude-limited sample in the Milky Way (\cite[e.g. Levesque \etal\ 2005]{levesque05}) contains objects of very different intrinsic brightness, including, for instance, stars of only $\sim6\:$M$_{\odot}$ (these objects are morphologically Ib supergiants, even if they are physically red giants). A direct comparison to, for example, a magnitude-limited sample in NGC~6822 (\cite[Levesque \& Massey 2012]{levesque12}), where only the brightest RSGs are included, would not be matching equivalent populations. Galaxy-wide surveys (\cite[e.g. Gonz\'alez-Fern\'andez \etal\ 2015]{carlos15}) mix together stars of different ages and, since many galaxies present metallicity gradients that may not readily be seen in projection, they should not be considered homogeneous populations.

\section{Clusters rich in red supergiants}

In recent years, helped by increasingly deep near- and mid-IR surveys, a number of clusters with large populations of RSGs have been found in the Milky Way. Because of their huge brightness in the IR, RSGs are easily seen even through very high extinction. Thanks to this, some clusters rich in RSGs -- or RSG clusters (RSGCs) -- have been detected behind very heavy obscuration. In these cases, determining cluster properties is not easy, as the IR spectra of RSGs do not provide information about their intrinsic brightness, and the accompanying population of less evolved stars may be quite faint. Table~\ref{tab1} lists known open clusters in the Milky Way with at least five RSG members, together with cluster parameters (in many cases, only estimates) and a rough estimation of the amount of interstellar extinction along their sightlines. The spatial distribution of known RSGCs on the Galactic Plane suggests that our current sample is far from complete, and so it is likely that more such clusters will be found in the near future.

\begin{table}[t]
  \begin{center}
  \caption{Open clusters in the Milky Way with at least 5 RSGs.
  \label{tab1}}
  \begin{tabular}{|l|c|c|c|c|}\hline 
{\bf Cluster} & {\bf \# RSGs} & {\bf Age} (Ma) & {\bf Mass estimate}& {\bf $A_{V}$} (mag) \\ 
&&& ($10^{4}\:$M$_{\odot}$)&\\\hline
NGC~884& 5 & 14 & 0.6\,--\,0.8& $<2$\\
NGC~7419& 5 & 14 & 0.6\,--\,0.9& $\approx5$\\
RSGC1& 13 (+1 YSG)& $\sim12$ & 2\,--\,4& $>20$\\
Stephenson~2& $\sim25$& 16\,--\,20 & $\sim5$& $\sim10$\\
RSGC3& $>10$&  16\,--\,20 & $\sim2$& $\sim13$\\
Alicante~7& $>8$&  16\,--\,20 & 1\,--\,2& $\sim13$\\
Alicante~10& $>8$&  16\,--\,20 & 1\,--\,2& 15\,--\,20\\
VdBH~222& 10 (+2 YSG)&  18\,--\,20 & $\sim1.5$& $\sim8$\\
Teutsch~85& 14 & $\sim25$& $\sim1$ & $\sim7$\\ \hline
  \end{tabular}
 \end{center}
\vspace{1mm}
 \scriptsize{
 {\it References:}\\
NGC~884: Currie \etal\ (2010); NGC~7419: Marco \& Negueruela (2013); RSGC1: Davies \etal\ (2008); Ste~2: Davies \etal\ (2007); RSGC3: Clark \etal\ (2009b); Al~7: Negueruela \etal\ (2011); Al~10: Gonz\'alez-Fern\'andez \& Negueruela (2012); VdBH~222: Marco \etal\ (2014); Te~85: Marco \etal\, in prep.}
\end{table}


RSGCs contain populations of RSGs that can be assumed homogeneous in metallicity and likely to have very similar masses. They are thus our best approximation to a single-age population. Stephenson~2 (Ste~2) contains a very large number of RSGs and lies behind obscuration that, although high, allows observations in the red and hence accurate spectral classification. These properties  probably make it our best template. Most RSGs in Ste~2 and in Milky Way clusters in general have spectral types in the range M0\,--\,M2. All the clusters in Table~\ref{tab1} contain some RSGs (at least, one) with later spectral types, which are systematically brighter. This probably explains the spectral distribution observed in the Galaxy (\cite[Levesque \etal\ 2005]{levesque05}), because late-M stars will be overrepresented in a magnitude-limited sample. In Ste~2, two very late RSGs (M6\,--\,M7) are associated with H$_{\textrm{2}}$O and SiO masers, and show very large mid-IR excesses, pointing to very heavy mass loss. This is also the case for the M7.5 RSG MY~Cep, in NGC~7419 (\cite[Verheyen \etal\ 2012]{verheyen12}). These stars are about 2~mag brighter than the faintest RSGs in their respective clusters. In Ste~2, there is an excellent correlation between spectral type, bolometric brightness and colours indicative of dust envelopes (\cite[Negueruela \etal\ 2013]{negueruela13}; \cite[Clark \etal\ 2014]{clark14}). In other clusters, the statistics are not so significant, but general trends are similar. This points very suggestively to an evolutionary sequence where stars reach the RSG phase with spectral types around M0 and subsequently increase their brightness and mass loss rates until reaching late-M types, characterised by heavy mass-loss and massive envelopes. The most extreme RSGs, such as VY Cam or VX~Sgr, are known to be rather massive ($>20\:$M$_{\odot}$), but the stars in Ste~2 are expected to have $\sim10\:$M$_{\odot}$, while those in NGC~7419 or the Per~OB1 association should be only slightly more massive. This suggests that very late spectral types and heavy mass loss might not be associated only with high mass, but perhaps also with an evolutionary phase of (most) RSGs. 

In addition to their role as laboratories, RSGCs can be used as probes for abundance determination in the inner Milky Way. RSGs are very bright targets that can be observed at high resolution, even if very distant. With the development of adequate diagnostics in the near-IR (e.g. \cite[Origlia \etal\ 2013, and references therein]{origlia13}), even the more obscured clusters can be used for this task, reaching regions where more traditional abundance tracers are not accessible. In this way, RSGCs open up a window to study the metallicity gradient in the central regions of the Milky Way. These twin roles as major astrophysical tools spur current efforts to detect and characterise new clusters rich in red supergiants.

\acknowledgements
I heartily thank all my collaborators, especially Amparo Marco and J. Simon Clark. Research partially supported by MinECO/FEDER under grant AYA2012-39364-C02-02.

\end{document}